\documentclass[a4paper,12pt]{article}

\usepackage[T2A]{fontenc}
\usepackage{amsmath}
\usepackage{amssymb}
\usepackage{xfrac}
\usepackage{cite}
\usepackage[unicode,linktocpage=true,plainpages=false,pdfpagelabels=false]{hyperref}

\newcommand{\sh}{\sinh}
\newcommand{\ch}{\cosh}

\begin{document}

\title{Explicit isometric embeddings of black holes geometry\\
with non-singular matter distribution}
\author{A.~D.~Kapustin\thanks{E-mail: sashakapusta96@gmail.com}\ \ and S.~A.~Paston\thanks{E-mail: pastonsergey@gmail.com}\\
{\it Saint Petersburg State University, Saint Petersburg, Russia}
}
\date{\vskip 15mm}
\maketitle

\begin{abstract}
The work is devoted to the construction of explicit embeddings for the metrics of the black holes, formed by nonsingular matter distribution.
One of the possible examples of such type of solutions is regular black hole.
Using the existing classification of minimal symmetric embeddings of the Schwarzschild metric as a base, we construct embeddings for regular black holes with de Sitter interior.
Another simple example is black hole, formed by collapsing homogeneous spherically symmetric cloud of dustlike matter.
We discuss embeddings for two variants of such black holes -- the one with the eternally existing horizon,
when dust ball never leaves the interior of the horizon, and another variant with the dynamically forming horizon.
\end{abstract}

\newpage

\section{Introduction}
It is known that arbitrary $d$-dimensional (pseudo)Riemannian space can be locally represented as a surface isometrically embedded into flat $N$-dimensional ambients space,
if $N \geqslant d(d+1)/2$ \cite{fridman61}. Then (pseudo)Riemannian space can be described by a set of embedding functions $y^a(x^\mu)$, and a metric can be considered as the \emph{induced} one
\begin{equation}
\label{emb} g_{\mu \nu} = (\partial_\mu y^a) \eta_{ab} (\partial_\nu y^b ),
\end{equation}
where $\eta_{ab}$ is a flat metric of ambient space (indices $\mu,\nu$ take $d$ values and $a,b$ take $N$ values).
Application of this approach to four-dimensional space-time allows us to obtain the modified
theory of gravity, which was proposed by Regge and Teitelboim in the work \cite{regge}
and later was studied under the names ``embedding theory'', ``geodetic brane gravity'' and ``embedding gravity'' in the number of works \cite{deser,pavsic85let,maia89,davkar,statja18,estabrook2009,statja25,faddeev}.
Having a simple geometric formulation, embedding gravity can be a fruitful modification of GR,
since the Einstein-Hilbert action, when written in terms of embedding functions, leads to Regge-Teitelboim equations, which are more general than Einstein equations.
It turns out that this theory contains additional degrees of freedom which can be used for explanation of the dark matter
\cite{davkar,estabrook2009,statja51}.
In addition, the use of flat space resolves some ideological problems \cite{carlip}, arising during the quantization of gravity in the framework of GR.

The visualizability of the embedding gravity approach turns out to be very useful for the description of complicated geometries, arising from Einstein equations. The study of explicit embeddings becomes especially fruitful  in case of geometries of black holes. The first explicit embedding of the Schwarzschild metric was obtained in 1921 \cite{kasner3}. Fronsdal embedding of the Schwarzschild metric, which is the most useful one for the study of global structure,  was obtained in 1959 \cite{frons}. It is smooth and covers the whole maximal analytical extension of the Schwarzschild solution, i.e. two areas inside the horizon and two copies of universes outside the black hole.
In addition to these, there are four other minimal embeddings (corresponding to a $6$-dimensional ambient space), as well as their classification based on the constructive method for obtaining embeddings of spaces with sufficiently rich symmetry. This method was proposed in the work \cite{statja27}. Along with embeddings of the Schwarzschild metric, embeddings of charged black holes, black holes with the presence of cosmological constant \cite{statja30} or rotating black holes \cite{kuzeev,kuzeevRN} were also studied.

All listed results correspond to the so-called eternal black holes,
which are formed by the matter concentrated in one point, i.e. has singular distribution.
Embeddings of black holes with nonsingular (except the singularity in the future or the past) distribution of matter weren't actively studied before, whereas they are of rather great physical interest. Such solutions have lower symmetry in general, so the task of construction of explicit embeddings for them is significantly harder. However, in some cases this problem can be overcome.

The first case is a consideration of the so-called regular spherically symmetric black holes \cite{dymnicova1, Dymnikova_2005}, which metric turns out to be static, and hence, has the same symmetry as Schwarzschild metric.
Staticity of the metric is achieved due to the fact that the properties of matter are not specified and the non-trivial equation of state, which provides the statics, is assumed for it. The construction of embeddings for metrics of such type is discussed in section 2.
The second case is a consideration of the movement of dustlike matter with a spherically symmetric distribution. Embeddings for this case were proposed in the work \cite{statja57} and discussed in section~3.

\section{Embeddings of regular black holes}
Spherically symmetric regular black hole is an exact solution of Einstein equations with some statically distributed matter, which leads to the lack of central singularity. In the works \cite{dymnicova1,Dymnikova_2005} regular black holes with de Sitter interior are discussed, i.e. in the asymptotics $r \to 0$ the metric of such black hole tends to the de Sitter metric
\begin{equation}
ds^2 = \left(1 - \frac{r^2}{r_c^2}\right) dt^2 - \left(1 - \frac{r^2}{r_c^2}\right)^{-1} dr^2 - r^2 d\Omega,
\end{equation}
where $d\Omega^2=d\theta^2+\sin^2\theta d\varphi^2$.
At the same time in the asymptotics $r \to \infty$ this metric tends to the Schwarzschild metric
\begin{equation}
ds^2 = \left(1 - \frac{R}{r}\right) dt^2 - \left(1 - \frac{R}{r}\right)^{-1} dr^2 - r^2 d\Omega,
\end{equation}
where $R$ is the Schwarzschild radius.
Note that it is possible to consider both charged in some way \cite{Dymnikova_2004} and rotating \cite{Dymnikova_2015} versions of regular black holes,
however, the search for corresponding embeddings is beyond the scope of this work.

The metric of a spherically symmetric regular black hole satisfying the described asymptotic conditions can be written in the general form
\begin{equation}
\label{metric}	ds^2 = g(r) dt^2 - \frac{dr^2}{g(r)} - r^2 d\Omega, \qquad
g(r) = 1 - \frac{1}{r}\int \limits_0^r \xi(\tilde r) \tilde r^2 d\tilde r,
\end{equation}
where $\xi(r)>0$ is proportional to the energy density. This function should have finite limit when $r \to 0$ in order to obtain the de Sitter asymptotics. On the other hand, when $r \to\infty$ the function $\xi(\tilde{r})$ should decrease fast enough for black hole mass to be finite. It can be easily seen that function $g(r)$ is bounded and satisfy the inequality $g(r) \leqslant 1$.

The metric \eqref{metric} has symmetry $SO(3) \times T^1$ (where $T^1$ denotes invariance with respect to time $t$ translations), the same as the Schwarzschild metric symmetry, and hence we can discuss possible embeddings for regular black holes, relying on the well-known classification of symmetric minimal embeddings of the Schwarzschild metric in a flat ambient space \cite{statja27}.
Two out of six known types of minimal embeddings for the Schwarzschild metric (elliptic and parabolic; see terminology used in \cite{statja40}) turn out to be non-global as for Schwarzschild case because they cannot be continued under the horizon.
The embedding of hyperbolic type, which is smooth for the Schwarzschild case (Fronsdal's embedding \cite{frons}),
turns out to be nonglobal for a regular black hole. The reason for this is the presence of two horizons.
The same situation was found in the work \cite{statja30} while trying to use this type of embeddings for the Reissner-Nordstrom metric.
The three remaining types of embeddings -- exponential, spiral and cubic
allow us to construct global embeddings for the metric \eqref{metric} of the spherically symmetric regular black hole.

The signature of all these embeddings is $(+,-,-,-,-,-)$ and for all of them a part of the components of the embedding functions is the same:
\begin{equation}
\label{spher}	y^3 = r \cos{\theta}, \qquad y^4 = r \sin{\theta} \cos{\varphi}, \qquad y^5 = r \sin{\theta} \sin{\varphi}.
\end{equation}

In case of the embedding of exponential type the ansatz for searching for the remaining components of the embedding function has the form
\begin{equation}
\label{dav}	y^0 = f(r) \sh{\big(\beta t + w(r)\big)}, \qquad y^1 = f(r) \ch{\big(\beta t + w(r)\big)}, \qquad y^2 = \gamma t + h(r).
\end{equation}
Substituting this ansatz into the equation on the embedding functions \eqref{emb} we obtain
\begin{gather}
\label{dav1}	f(r) = \frac{1}{\beta}\sqrt{g(r)+\gamma^2}, \qquad h^\prime (r) = \frac{1}{\gamma \beta}\big(g(r) + \gamma^2\big) w^\prime (r), \\
\label{dav2}	w^\prime(r) = \frac{\gamma}{g(r)\sqrt{g(r)+\gamma^2}}
\sqrt{\big(1-g(r)\big)\left[ \beta^2 - \frac{g(r) g^\prime{}^2(r)}{4\big(1-g(r)\big) \big(g(r)+\gamma^2\big)} \right]}.
\end{gather}
Since function $g(r)$ is bounded, $\gamma$ can be chosen so that the expression under the square root in the formula \eqref{dav1} is positive.
In the formula \eqref{dav2} the expression under the root can be made positive by the choice of parameter $\beta$,
since $g(r) \leqslant 1$ (see above), and the fraction subtracted from $\beta^2$
is bounded, since, in accordance with \eqref{metric},
when $r \to 0$ we have $(1 - g(r))\sim r^2$, $g^\prime(r) \sim r$, and when $r \to \infty$ we have $\big(1 - g(r)\big)\sim \frac{1}{r}$, $g^\prime(r) \sim \frac{1}{r^2}$.
Since $g(r)$ turns into zero on the horizons, functions $h(r)$ and $w(r)$, in accordance with \eqref{dav1} and \eqref{dav2},
have logarithmic singularities. However, this singularities turn out to be coordinate ones,
as well as in the case of the Schwarzschild metric so they can be removed by moving from Schwarzschild time $t$
to a new time $t^\prime = t + h(r)/\gamma$, see \cite{statja27} for more details.

In case of the embedding of spiral type the ansatz for searching $y^0,y^2,y^2$ has the form
\begin{equation}
\label{asymp}	y^0 =\gamma t + h(r), \qquad y^1 = f(r) \sin{\big(\alpha t + w(r)\big)}, \qquad y^2 = f(r) \cos{\big(\alpha t + w(r)\big)}.
\end{equation}
Substituting this ansatz into the equation \eqref{emb} we obtain
\begin{gather}
\label{dav1a}	f(r) = \frac{1}{\alpha}\sqrt{\gamma^2 - g(r)}, \qquad h^\prime (r) = \frac{1}{\gamma \alpha}\big(\gamma^2 - g(r)\big) w^\prime (r), \\
\label{dav2a}	w^\prime(r) = \frac{\gamma}{g(r) \sqrt{\gamma^2 - g(r)}}\sqrt{\big(1-g(r)\big)\left[ \alpha^2 - \frac{g(r) g^\prime{}^2(r)}{4 \big(1-g(r)\big)\big(\gamma^2 - g(r)\big)} \right]}.
\end{gather}
We can show that all functions under square roots can be made positive by the choice of the parameters $\alpha$ and $\gamma$ using exactly the same reasoning as in the analysis of relations \eqref{dav1},\eqref{dav2}.
It should be noted that, as in the case of the Schwarzschild metric, the choice of parameter $\gamma = 1$
makes this embedding of regular black hole asymptotically ($r \to \infty$) flat, but in this case the smoothness disappears when $r\to0$. As well as for the previous embedding, coordinate singularities in \eqref{asymp}
can be removed by change of variable $t^\prime = t + h(r)/\gamma$.

Finally, let's consider the embedding of cubic type.
In this case it is convenient to use lightlike coordinates $y^\pm = (y^0 \pm y^1)/\sqrt{2}$ in the ambient space.
Then the ansatz for searching embedding functions has the form
\begin{equation}
\label{cub}	y^+ = h_1(r) + \gamma t \, h_2(r) + \frac{\gamma^2 t^2}{2} h_3(r) + \frac{\nu \, \gamma^3 t^3}{6}, \quad y^2 = h_2(r) + \gamma t \, h_3(r) + \frac{\nu \, \gamma^2 t^2}{2}, \quad y^- = h_3(r) + \nu \, \gamma t.
\end{equation}
The substitution of this ansatz into equations \eqref{emb} leads to expressions
\begin{gather}
h_1{}^\prime(r) = \frac{1}{\nu} \left( h_2^\prime(r) h_3(r) - h_3^\prime(r) h_2(r) \right), \qquad h_2(r) = \frac{1}{2 \nu \gamma^2}\big( g(r) + \gamma^2 h_3^2(r) \big), \\
h_3{}^\prime(r) = \frac{1}{\gamma g(r)} \sqrt{ \big(1 - g(r)\big) \left[ \nu^2 \gamma^4 - \frac{g(r)g^\prime{}^2(r)}{4 \big(1 - g(r)\big)} \right]}.
\end{gather}
One can make the function under the root positive by choosing the parameters $\nu$ and $\gamma$, because the fraction subtracting from $\nu^2 \gamma^4$ is bounded and differs from similar fractions in the expressions \eqref{dav2} and \eqref{dav2a} by multiplication by a bounded function.
As well as in the previous cases,  coordinate singularities in \eqref{cub} can be removed by the change $t^\prime = t + h_3(r)/(\nu\gamma)$.

\section{Embedding of black holes with dust-like matter}
In the consideration of the collapse of dust-like matter, the natural coordinate choice is the comoving coordinate frame
with coordinates $(\tau, \chi, \theta, \varphi)$.
The metric, corresponding to the solution of the Einstein equations with dustlike matter, has the form
\begin{equation}
\label{metric2} d s^2 = d \tau^2 - \frac{(r'\big(\tau, \chi)\big)^2}{1+f(\chi)} d\chi^2 - r^2(\tau, \chi) d \Omega^2,
\end{equation}
where prime denotes derivatives w.r.t. variable $\chi$.
Here $f(\chi)$ is related to the matter distribution and
$r(\tau, \chi)$ is a function, whose explicit form (see details in \cite{statja57}) is found by solving Einstein equations and
all that solutions belong to one of three classes, corresponding to the classification of the space-time as bound, marginally
bound or unbound \cite{joshi}.

At first let us consider a case similar to the studied in the previous section, because in both cases the horizon exists all the time. This case corresponds to the motion of matter, in which a homogeneous dust ball flies out from the white hole singularity, expands, reaching the Schwarzschild radius, and then falls into the black hole singularity.
The metric outside of the dust ball (and therefore beyond the horizon, because the matter is not there) corresponds to the Schwarzschild solution.

For this case, it is possible to construct a smooth six-dimensional embedding with the signature $(+, -, -, -, -, -)$, based on the junction of the mentioned above Fronsdal embedding for the Schwarzschild metric and some modification of the well-known $5$-dimensional embedding $y_f (x)$ \cite{robertson1933} for the metric of the closed Friedmann model, which has the form
\begin{equation}
\begin{gathered}
y^0_f = f(\tau), \quad y^1_f = a(\tau) \cos(\chi), \\
y^2_f = a(\tau) \sin(\chi) \cos{\theta}, \quad y^3_f = a(\tau) \sin(\chi) \sin{\theta} \cos{\varphi}, \quad y^4_f = a(\tau) \sin(\chi) \sin{\theta} \sin{\varphi}.
\end{gathered}
\end{equation}
The Fronsdal embedding $y_s(x)$ \cite{frons} under the horizon (this is where the junction of solutions should take place,
because in this case the matter never leaves the horizon)
has the form
\begin{equation}
\label{frons}	\begin{gathered}
y^0_s = \pm 2R \sqrt{\frac{R}{r} - 1} \ch{\left( \frac{t}{2R} \right)}, \quad y^1_s = 2R \sqrt{\frac{R}{r} - 1} \sh{\left( \frac{t}{2R} \right)}, \quad y^2_s =R \, q \left( \frac{r}{R} \right), \\
y^3_s = r \cos{\theta}, \quad y^4_s = r \sin{\theta} \cos{\varphi}, \quad y^5_s = r \sin{\theta} \sin{\varphi},
\end{gathered}
\end{equation}
where $q(x) = \int dx\sqrt{\frac{1}{x^3} + \frac{1}{x^2} + \frac{1}{x}}$.
It turns out that if junction surface is $\chi = \chi_0 = \pi/2$ then on this surface
blocks $y^2_f, y^3_f, y^4_f$ and $y^3_s, y^4_s, y^5_s$ coincide with each other; at the same time components $y^1_f$ and $y^1_s$
turn to zero (see details in \cite{statja57}) and then also coincide with each other.
The remaining step is to modify the $5$-dimensional embedding of the Friedman metric by replacing the component $y^0_f$ by two, which can be joined with $y^0_s$ and $y^2_s$.
The embedding constructed
by this method
can be written in the most convenient way using two coordinates of the ambient space $y^0, y^1$ instead of the coordinates $\tau, \chi$, which leads to the appearance of a coordinate singularity. For the interior of the matter cloud:
\begin{equation}
\begin{gathered}
y^2 = R \, q \left( \frac{\tilde r(y^0,y^1)}{R} \right), \quad
y^3 = \tilde r(y^0,y^1) \cos{\theta}, \\
y^4 = \tilde r(y^0,y^1) \sin{\theta} \cos{\varphi},\quad
y^5 = \tilde r(y^0,y^1) \sin{\theta} \sin{\varphi},
\end{gathered}
\end{equation}
and for the outside area:
\begin{equation}
\begin{gathered}
y^2 = R \, q
\left( \frac{\tilde r(y^0,0)}{R} \right),\quad
y^3 = \sqrt{\tilde r(y^0,0)^2-{y^1}^2} \cos{\theta}, \\
y^4 = \sqrt{\tilde r(y^0,0)^2-{y^1}^2} \sin{\theta}\cos{\varphi},\quad
y^5 = \sqrt{\tilde r(y^0,0)^2-{y^1}^2} \sin{\theta}\sin{\varphi},
\end{gathered}
\end{equation}
where $\tilde r(y^0,y^1)=\frac{4R^3}{{y^0}^2-{y^1}^2+4R^2}$.
The resulting embedding is smooth and global.

Now let us consider a physically more interesting case when horizon forms dynamically. Then at the initial moment the homogeneous dust ball has a radius greater than its gravitational radius $R$.
In this case, it is convenient to use the coordinate $r$ instead of the coordinate $\chi$ leading to a coordinate singularity, as already mentioned.
If we perform this change of coordinates for the case of marginally bound space-time, when $f(\chi)=0$ (see details in \cite{statja57}),
the metric \eqref{metric2} takes the form
\begin{equation}
\label{metric3}
d s^2 = \left(1-\frac{F(\tau, r)}{r} \right)d\tau^2 - 2 \sqrt{\frac{F(\tau, r)}{r}}dr d\tau - dr^2 - r^2 d\Omega^2,
\end{equation}
where function $F(\tau, r) = \min{ \left( \frac{4r^3}{9\tau^2}, R \right)}$ corresponds to the homogeneous matter distribution.

If we choose \eqref{spher} as three components $y^3,y^4,y^5$, and perform a change of coordinates
$p = r^3/\tau^2$, $t = \tau^{\sfrac{2}{3}}$, then the problem reduces to the construction of an embedding of a two-dimensional metric polynomially dependent on  the new time $t$
\begin{equation}
\label{eq10}	d s^2 = \left(\frac{9}{4} t +3 p^{\sfrac{1}{6}} \sqrt{\bar F(p)}-\frac{9}{4}p^{-\sfrac{1}{3}} \bar F(p) \right) dt^2 +
t\, p^{-\sfrac{5}{6}}\sqrt{\bar F(p)} \, dp\, dt
\end{equation}
where $\bar F(p)=\min{ \left( \frac{4}{9}p, R \right)}$.
Solving it, we obtain a $7$-dimensional embedding \cite{statja57}
\begin{equation}
\label{sp10}
\begin{gathered}
y^0 = \tau^2 + \frac{9}{16}\tau^{\sfrac{4}{3}}+\frac{1}{2}\tau^{\sfrac{2}{3}} \left( u \left( \frac{r^3}{\tau^2}\right)+1\right), \quad
y^1 = \tau^2 + \frac{9}{16}\tau^{\sfrac{4}{3}}+\frac{1}{2}\tau^{\sfrac{2}{3}} \left( u \left( \frac{r^3}{\tau^2}\right)-1 \right), \\
y^2 = w\left(\frac{r^3}{\tau^2}\right), \qquad
y^6 = \sqrt{\frac{3}{2}}\tau^{\sfrac{4}{3}}-w \left( \frac{r^3}{\tau^2}\right),
\end{gathered}
\end{equation}
with signature $(+,-,+,-,-,-,-)$,
where functions $u$ and $w$ are defined by the expressions
\begin{equation}
\begin{gathered}
u(p) = 3 p^{\sfrac{1}{6}} \sqrt{\bar F(p)}-\frac{9}{4}p^{-\sfrac{1}{3}} \bar F(p), \\
w(p) = \frac{1}{\sqrt{6}}\theta \left( p-\frac{9}{4} R \right)
\left( \left( \frac{9}{4} R \right)^{\sfrac{1}{2}}p^{\sfrac{1}{6}}+\frac{9}{8} R p^{-\sfrac{1}{6}}-\frac{3}{2}\left(\frac{9}{4} R \right)^{\sfrac{2}{3}}\right).
\end{gathered}
\end{equation}
Time $\tau$ is considered to be negative and its zero point corresponds to the simultaneous fall of all particles of dustlike matter on a singularity.
Like all of the above, this embedding is smooth and global, however, in contrast to others, this embedding corresponds to the process of dynamic formation of the black hole horizon.

{\bf Acknowledgements}
The work is supported by RFBR Grant No.~20-01-00081.


\end{document}